\RequirePackage{fix-cm}
\documentclass[twocolumn]{svjour3}          % twocolumn
\smartqed  % flush right qed marks, e.g. at end of proof

\usepackage[square]{natbib}

\usepackage{graphicx}
\usepackage{siunitx}
\usepackage[draft]{listofsymbols}
\usepackage{amsmath,amssymb}
\usepackage{tabularx}
\usepackage{booktabs}
\usepackage{shellesc}
\usepackage{tikz}
\usetikzlibrary{external}
\usepackage{pgfplots} 
\usepgfplotslibrary{external}
\usepackage{pgfgantt}
\usepackage{pdflscape}
\usepackage[export]{adjustbox}
\usepackage[permil]{overpic}
\usepackage{subfig}
\pgfplotsset{compat=newest} 
\pgfplotsset{plot coordinates/math parser=false}
\usepackage{verbatim} 
\usepackage{ marvosym }
\usepackage[normalem]{ulem}
\usepackage{dblfloatfix}    % To enable figures at the bottom of page

\usetikzlibrary{shapes.arrows}

\usepackage[colorlinks,linkcolor=blue,citecolor=blue,urlcolor=blue,bookmarks=false,hypertexnames=true]{hyperref}

\usepackage[switch,columnwise]{lineno}
\usepackage{dblfloatfix}

\newcommand{\uproman}[1]{\uppercase\expandafter{\romannumeral#1}}

\begin{document}

\sloppy

\title{Two-Color LIF investigation of mixing during droplet impact onto a thin liquid film}
%\subtitle{LIF measurements to study the mixing process during droplet impact onto thin liquid film}

\author{Hatim Ennayar \and Jeanette Hussong%etc.
}
 
 \institute{
 {Hatim Ennayar} (\Letter)  \and Jeanette Hussong \at
             Institute for Fluid Mechanics and Aerodynamics\\
             Technische Universität Darmstadt 
             \\ Darmstadt, Germany \\ 
              \email{ennayar@sla.tu-darmstadt.de}            
                               }

\date{Received: date / Accepted: date}

\maketitle

\begin{abstract}
A two-color laser-induced fluorescence (2C-LIF) technique is presented for investigating droplet impact on thin liquid films, enabling simultaneous, spatially and temporally resolved measurements of film thickness and scalar concentration. The method is applied to water droplets impacting thin films over a range of Reynolds numbers, Weber numbers, and dimensionless film thicknesses, providing direct access to early-time mixing processes during impact. To quantify scalar transport within the liquid film, the reconstructed concentration fields are evaluated using a coefficient-of-variation (CV) approach, providing a quantitative measure of mixture homogeneity. This enables identification of the transition from inertia-dominated convective transport to diffusion-controlled mixing. Based on this analysis, an empirical correlation describing the evolution of the CV as a function of Reynolds number and film thickness is formulated. Finally, the applicability of the method is demonstrated for binary ethanol–water films, where additional transport mechanisms influence and modify the mixing dynamics.
\end{abstract}

\section{Introduction}
\label{sec:1}
Droplet impact on thin liquid films is a fundamental process encountered in nature and in a wide range of industrial applications. Despite its apparent simplicity, the interaction between an impacting droplet and a wetted surface gives rise to complex and highly transient fluid dynamic behavior. In many industrial processes, controlled droplet–film interactions are essential. A prominent example is spray coating, used in applications such as automotive painting (\citealt{akafuah2016evolution}), protective coatings (\citealt{fotovvati2019coating}), pharmaceuticals (\citealt{bose2007solventless}) and optical manufacturing (\citealt{xiao2025spray}). In these systems, the mixing between deposited droplets and the underlying film determines coating uniformity and functional performance, while insufficient mixing can lead to defects and spatial variations in material properties (\citealt{gutoff2006coating,fitzsimons2015coating}). The sensitivity of these processes arises from the underlying flow configuration, in which the dynamics are governed not only by the impacting droplet but also by the finite thickness of the liquid layer. The impact process is commonly characterized by the Reynolds number $Re=U D/\nu$ and the Weber number $We=\rho U^2 D/\sigma$, where $U$ is the impact velocity, $D$ the droplet diameter, $\rho$ the liquid density, $\nu$ the kinematic viscosity, and $\sigma$ the surface tension. In addition, the dimensionless film thickness $\delta=h/D$, defined by the ratio of the film thickness to the droplet diameter, introduces a geometric control parameter that governs the influence of the liquid layer and the solid boundary (\citealt{tropea1999impact,geppert2019experimental}). 

While extensive research has established regime maps describing impact outcomes such as deposition, crown formation, and splashing (\citealt{yarin2006drop,cossali1997impact}), these primarily describe the evolution of the free surface. In contrast, the internal transport processes governing the mixing between droplet and film liquid remain insufficiently quantified, particularly during the early, inertia-dominated stages of impact. 

Several studies have investigated mixing during droplet impact using dye-based visualization techniques (\citealt{ersoy2019capillary,ersoy2020phenomenological,khan2020droplet,parmentier2023drop}). For example, \citealt{ersoy2020phenomenological} examined the mixing of water droplets impacting water films using methylene blue as a tracer for film thickness ratios ranging from $\delta = 0.045$ to $\delta = 1.768$. To quantify the degree of mixing, they introduced a metric based on grayscale image analysis, in which a thresholded intensity signal identifies regions affected by the dyed droplet. The mixing efficiency was then defined as the ratio of the film area covered by dye to the total imaged film area. This approach, also adopted in \citealt{ersoy2019capillary} and in \citealt{khan2020droplet}, infers mixing from thresholded grayscale intensity fields obtained from top-view dye visualizations. While such methods provide qualitative estimates of the spatial extent of mixing, they do not resolve local concentration fields within the liquid film. In particular, the thresholding approach yields essentially binary information on the presence of dyed liquid and does not provide depth-resolved concentration measurements. A more advanced approach was proposed by \citealt{parmentier2023drop}, who developed a colorimetry-based method using two dyes to distinguish liquid originating from the drop and from the film. By calibrating RGB intensity values against known dye concentrations, they reconstructed spatial maps of the post-impact droplet liquid distribution. However, their analysis was performed at a single time instant approximately one second after impact, once convective motion had largely subsided. Consequently, although spatial concentration distributions were obtained, the transient early-time convective transport processes that dominate mixing during the impact itself were not resolved.

This indicates a clear research gap in the quantitative characterization of the transient, inertia-driven mixing processes occurring during the early stages of droplet impact on thin liquid films. In contrast to colorimetry-based techniques relying on absorbance, fluorescence-based diagnostics offer significantly higher sensitivity for detecting small concentration variations in liquid flows (\citealt{rasheed2024advancements,alqarni2023colorimetric}). In the present work, fluorescence-based laser-induced fluorescence (LIF) technique is therefore employed to quantitatively reconstruct temporally and spatially resolved concentration fields in thin liquid films during droplet impact. 

Laser-induced fluorescence has become a widely used method for investigating scalar transport and mixing in liquid flows, including thin-film configurations. Following its first quantitative application by \citealt{owen1976simultaneous} for measuring instantaneous velocity and concentration in turbulent mixing flows, LIF has been widely adopted for scalar transport measurements in fluid mechanics (\citealt{koochesfahani1985laser,walker1987fluorescence,arcoumanis1990use}). While these studies relied on light-sheet illumination, volume illumination approaches enable access to the scalar distribution across the full film depth and are therefore particularly suited for thin-film applications. In addition to concentration measurements, LIF has also been employed to reconstruct film thickness variations (\citealt{ennayar2023lif,hann2016study}), which is of particular relevance in droplet impact experiments where the fluorescence signal depends on both dye concentration and film thickness as will be explained in Sec. \ref{sec:3}. In its conventional implementation, LIF relies on a single fluorescent dye and a single detection wavelength band, such that the recorded signal represents a coupled response to multiple scalar quantities. As a result, variations in concentration and film thickness cannot be independently resolved from a single measurement. This limitation was also present in our previous work (\citealt{ennayar2023lif}), where single-color LIF was applied to investigate droplet mixing dynamics on quiescent thin films. In that study, concentration and film thickness had to be determined through separate experimental procedures, increasing both the experimental effort and the susceptibility to systematic uncertainties. Two-color laser-induced fluorescence (2C-LIF) overcomes this by decoupling of concentration and thickness effects. 2C-LIF has been developed and applied for liquid phase thermometry (\citealt{collignon2021temperature,mishra2020investigation,koegl2024simultaneous,mishra2016thermometry}) or fluid mixing studies (\citealt{koegl2020novel,zhang2014simultaneous,mohri2011imaging}). More recently, 2C-LIF has been used to simultaneously capture temperature and concentration fields of binary mixture droplets during jet breakup (\citealt{ulrich2024two}). 

Building on these developments, the present study extends our previous single-color LIF methodology (\citealt{ennayar2023lif}) to a two-color LIF framework for droplet impact on thin liquid films. The first objective is to establish and validate a 2C-LIF approach that enables the simultaneous reconstruction of local film thickness and scalar concentration from a single experiment, thereby overcoming the sequential measurement procedure required in the earlier single-color implementation. The second objective is to apply this technique to quantify the transient mixing dynamics during droplet impact on thin films over a range of Reynolds numbers, Weber numbers, and dimensionless film thicknesses. To the authors’ knowledge, this is the first parametric experimental study to resolve and quantify early-time mixing during droplet impact on thin liquid films using spatially and temporally resolved concentration fields. Finally, the applicability of the approach is extended beyond single-component systems by demonstrating measurements in binary ethanol–water films, in which additional composition-dependent transport mechanisms arise.

\section{Experimental setup}
\label{sec:2}
\begin{figure}
\centering
\includegraphics[scale=0.75]{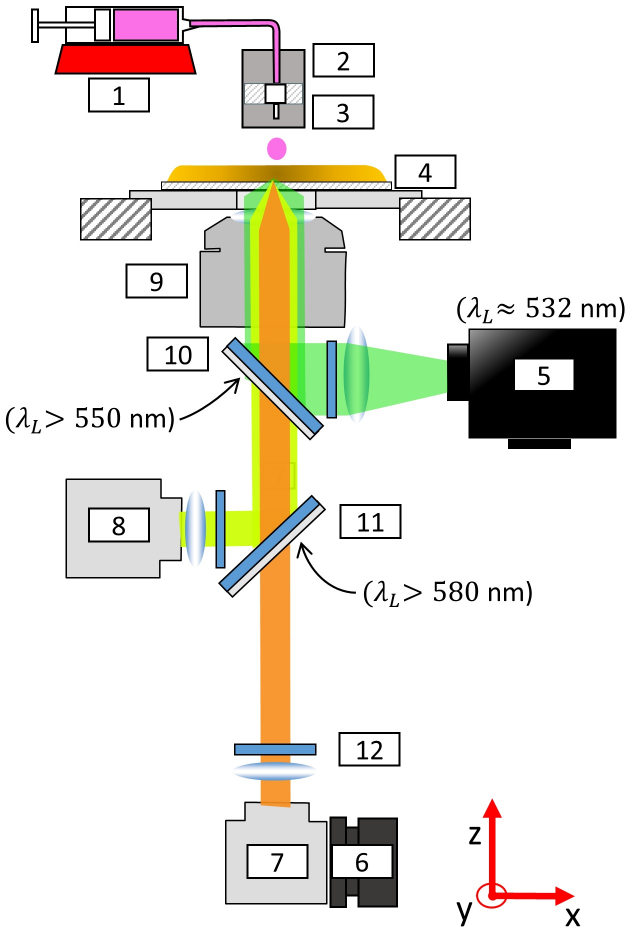}
\caption{Schematic illustration of bottom view setup: (1) Syringe pump, (2) z-Traverse, (3) Cannula, (4) Thin liquid film on FTO glass substrate, (5) High power LED, (6) x,y,z-Traverse, (7) HS-Camera, (8) HS-Camera, (9) Microscope Objective, (10) Dichroic mirror (\textbf{$\lambda_L>\SI{550}{\nano\meter}$}) with bandpass filter (\textbf{$\SI{532}{\nano\meter}$}), (11) Dichroic mirror (\textbf{$\lambda_L>\SI{580}{\nano\meter}$}) with bandpass filter (\textbf{$\SI{561}{\nano\meter}$}), (12) Bandpass filter (\textbf{$\SI{624}{\nano\meter}$}).}
\label{fig:1}
\end{figure}

The experimental setup builds upon a previously established configuration originally designed for single-color LIF studies (\citealt{ennayar2023lif}). In the present study, this setup was extended to a two-color LIF approach, which enables simultaneous measurements of film thickness and species concentration. A schematic of the optical set-up is shown in Figure \ref{fig:1}.

Droplets are generated by dispensing water through a blunt-tip needle (Braun GmbH) connected to a syringe (Braun GmbH) that is driven by a syringe pump (Aladdin AL-1010, WPI). Mounting the needle on a motorized traverse provided accurate control of the release height and thus of the impact velocity up to of 2 m/s. Water droplets of diameters of $D = 2.225 \pm \SI{0.025}{\milli\meter}$ were used. Depending on impact conditions, Reynolds numbers of $1800$, $3000$ and $4200$ were investigated, corresponding to Weber numbers of $24$, $54$ and $110$. Impacts took place on a glass substrate ($50 \times 50$ mm$^2$, Sigma-Aldrich) that was thoroughly cleaned in an ultrasonic bath using isopropanol and subsequently rinsed with deionized water. Prior to each experiment, a thin liquid film was spread on the substrate, with the contact line pinned at the four edges to ensure reproducibility. Film thicknesses of $200$, $500$ and $\SI{800}{\micro\meter}$ were investigated, corresponding to dimensionless film thicknesses $\delta=0.09$, $0.22$ and $0.36$, respectively. Thickness measurements prior to the impact were obtained using a chromatic-confocal point sensor (confocalDT IFS2407-0.8, Micro-Epsilon), which provides an accuracy of $\pm \SI{0.4}{\micro\meter}$. Experiments were performed with water films as well as binary water-ethanol mixtures at 10\% and 30\% ethanol mass fractions. The corresponding liquid properties are summarized in Table \ref{tab:1}

The optical arrangement consisted of a custom-built microscope designed to enable high-speed, coaxial fluorescence imaging of the droplet impact region. A continuous-wave high-power green LED ($\lambda \approx \SI{532}{\nano\meter}$, ILA iLA.LPS v3) was used for coaxial illumination. The fluorescence signal was collected through a $1\times$ microscope objective (Infinity Photo-Optical IF-3) and split into two spectral bands of $561 \pm 10.5$ nm and $624 \pm 23$ nm using dichroic mirrors and band-pass filters mounted in filter cubes (Thorlabs). The two signals were recorded simultaneously by synchronized high-speed CMOS cameras operating at 6000 fps: a Phantom T3610 (12-bit, $1200 \times 800$ pixels) and a Phantom T1340 (12-bit, $2048 \times 1952$ pixels). The recorded signals were subsequently combined by once superposing images and once taking their intensity ratio. In this way, two observables can be used for evaluating liquid film thickness and concentration. The detailed calibration and evaluation procedure is given in Sec. \ref{sec:3}. The optical set-up is mounted on a three-axis motorized precision stage capable of moving in increments of $1.25~\mathrm{\mu m}$.

\begin{table}
\centering
\begin{tabularx}{\columnwidth}{lXXX}
\hline
Property & Water & 10\% EtOH & 30\% EtOH \\
\hline
Density $\rho$ (kg/m$^3$) & 998 & 982 & 954 \\
Kinematic viscosity $\nu$ (mm$^2$/s) & 1.004 & 1.62 & 2.64 \\
Surface tension $\sigma$ (N/m) & 0.0723 & 0.0481 & 0.0335 \\
\hline
\end{tabularx}
\caption{Material properties of the liquid used in thin films.}
\label{tab:1}
\end{table}

Two different fluorescent dyes were added to the liquid film, while only one of them was present in the impacting droplet. In this configuration, the fluorescence signals detected in the two spectral channels provide independent information that enables the simultaneous reconstruction of film thickness and species concentration. Although the optical response is nonlinear as will be seen in Sec. \ref{sec:3}, this approach establishes a determined two-parameter relationship between the measured intensities and the physical quantities of interest, simplifying calibration, lowering the experimental effort and avoiding the risk of non-unique inversion.

RhB and Rhodamine~6G (Rh6G, Carl Roth) were employed as tracers. RhB was added at $7.5 \times 10^{-5}\mathrm{M}$ to both droplet and film, while Rh6G was introduced only into the film at the same concentration. The selection of RhB and Rh6G is motivated by their compatibility with a common excitation source and their favorable spectral separation. Both dyes can be efficiently excited using the same green LED illumination, since Rh6G exhibits an absorption maximum at approximately $\SI{530}{\nano\meter}$ and an emission maximum near $\SI{565}{\nano\meter}$ (\citealt{chapman2018rhodamine}), which is spectrally distinct from the emission of RhB. To avoid contributions from the emission of Rh6G in the RhB detection channel, the emission filter used in the single-color LIF configuration (centered at 592 nm) was replaced in the two-color setup by a filter centered at 624 nm. This choice suppresses contributions from the tail of the Rh6G emission spectrum in the RhB detection channel.

The labeling strategy described above is applicable for experiments conducted with pure water films. For ethanol-water mixture films, however, an additional dependency arises from the solvent-sensitive fluorescence behavior of Rh6G. While Rh6G remains mostly monomeric in ethanol-rich environments, higher water fraction can promote aggregation, thereby altering the absorption/emission characteristics and reducing fluorescence yield (\citealt{chapman2018rhodamine}). As a consequence, the measured intensity becomes a function not only of film thickness and dye concentration but also of the local ethanol fraction. Under these conditions, variation in the dye concentration during droplet impact analogously to the water case would introduce a third independent variable, which would require a three-color LIF approach. To circumvent this limitation, Rh6G was added to both the droplet and the film in the ethanol-water experiments, while maintaining fixed concentrations of both dyes in both the liquid film and the droplet. In this configuration, variations in the fluorescence response are primarily governed by liquid film thickness and ethanol concentration. Calibration samples with varying ethanol fraction were subsequently prepared to establish the relationship between the sum and ratio of the fluorescence intensities recorded in the two detection channels and the corresponding physical quantities, namely the local film thickness and ethanol concentration. The calibration procedure is described in detail in the next section.

\section{Method and calibration procedure}
\label{sec:3}
With the experimental configuration established, the following section presents the procedure for the 2C-LIF measurements. To quantitatively resolve local film thickness and dye concentration during droplet impact, a dedicated calibration procedure was developed that accounts for the parameters influencing the recorded fluorescence intensity. The detected fluorescence signal, $I_{f}$, can be described in simplified form as (\citealt{guilbault1973}):
\begin{equation}
	I_{f} = A I_{0} \Phi \epsilon L C ,
	\label{eq:1}
\end{equation}
where $I_{0}$ denotes the excitation light intensity, $\Phi$ the fluorescence quantum yield, $\epsilon$ the molar absorptivity, $L$ the optical path length (here corresponding to film thickness), and $C$ the dye concentration. The proportionality constant $A$ represents the fraction of emitted light collected by the imaging system.

The fluorescence response described in Eq. (\ref{eq:1}) is assumed to vary linearly with concentration within concentration ranges and illuminated volume depths that are specific to the fluorescent dye used. For Rhodamine B, this linear relationship is generally maintained at concentrations below $\SI{0.08}{\milli\gram\per\litre}$ (\citealt{arcoumanis1990use}). However, \citealt{wigger2016quantitative} showed that noticeable deviations from linearity occur when the product $LC$ exceeds about $\SI{5}{\micro\meter\:\milli\mole\per\litre}$. In the present experiments, concentrations up to $7.5 \times 10^{-5}$ M (approximately $\SI{36}{\milli\gram\per\litre}$) and film thicknesses of several hundred micrometres were generated. Under these conditions, the response departs from strict linearity, requiring multidimensional calibration to accurately map the measured fluorescence signals to local film thickness and dye concentration. Spatial variations in illumination across the field of view were also compensated by including pixel-wise coordinates in the calibration.

\begin{table*}
	\centering
	\begin{tabularx}{\textwidth}{X X X X X}
		\toprule
		\textbf{Film composition} &
		\textbf{Droplet dye} &
		\textbf{Film dye} &
		\textbf{Varied in calibration} &
		\textbf{Fixed in calibration} \\
		\midrule
		Pure water &
		RhB &
		RhB + Rh6G &
		$h$, $C_{\mathrm{Rh6G}}$ &
		$C_{\mathrm{RhB}}$ \\
		Ethanol–water mixtures &
		RhB + Rh6G &
		RhB + Rh6G &
		$h$, $\phi_{\mathrm{EtOH}}$ &
		$C_{\mathrm{RhB}},\,C_{\mathrm{Rh6G}}$ \\
		\bottomrule
	\end{tabularx}
	\caption{Fluorescent labeling and calibration strategies employed for different liquid film compositions. Here $h$ denotes the liquid film thickness, $C$ the concentration of the corresponding dye and $\phi_{\mathrm{EtOH}}$ the ethanol mass fraction.}
	\label{tab:2}
\end{table*}

To quantitatively resolve local liquid film thickness and dye or ethanol concentration during droplet impact, a dedicated calibration procedure was developed that accounts for the parameters influencing the recorded fluorescence intensity. Fluorescence was recorded simultaneously in two spectral channel. The measured intensities in these channels, denoted by $I_{1}(x,y)$ and $I_{2}(x,y)$, were combined to form the total intensity
\begin{equation}
	S(x,y) = I_{1}(x,y) + I_{2}(x,y),
	\label{eq:2}
\end{equation}
and the intensity ratio
\begin{equation}
	R(x,y) = \frac{I_{2}(x,y)}{I_{1}(x,y)}.
	\label{eq:3}
\end{equation}

Prior to forming $(S,R)$, the two camera images were geometrically aligned and filtered as described below.

The relationship the measured observables $(S,R)$ and the physical quantities of interest, namely the local film thickness $h$ and dye or ethanol concentration $C$, was determined independently for each pixel using polynomial regression. This pixel-wise calibration strategy accounts for spatial nonuniformities such as illumination gradients and optical vignetting across the field of view (\citealt{likar2000retrospective,marty2007blank}).

Calibration data were acquired on the same substrate and under identical optical alignment as in the impact experiments. Water films of known thicknesses ranging from $h = 0$ to $1000~\si{\micro\meter}$ in increments of $\SI{100}{\micro\meter}$ were prepared. RhB was maintained at a constant concentration of $7.5\times10^{-5}~\mathrm{M}$, whereas the Rh6G concentration was varied between $0$ and $7.5\times10^{-5}~\mathrm{M}$. For the ethanol–water mixtures, the same thickness range was covered while the ethanol fraction was varied between 0 and 30\% w/w. In this case, both RhB and Rh6G were present in the film at fixed concentrations. An overview of the fluorescent labeling strategies and the parameters varied and held fixed during calibration for each film composition is provided in Table~\ref{tab:2}.

For each combination of $(h,C)$, synchronized image pairs $(I_1,I_2)$ were recorded, and the corresponding quantities $(S,R)$ were computed. Since the two high-speed cameras differed in sensor and pixel sizes, the recorded images were geometrically aligned using bicubic interpolation onto a common coordinate grid with an unified pixel size of $\SI{16.42}{\micro\meter}$. The resulting field of view covered $15.5\times15.5~\si{\milli\meter\squared}$. Prior to forming $(S,R)$, Noise reduction was applied to each channel using a median filter with a $5\times5$ kernel.

For each pixel, two independent nonlinear regression models were fitted,
\begin{equation}
	h = f_h(S,R), \qquad C = f_C(S,R),
	\label{eq:4}
\end{equation}
where $f_h$ and $f_C$ are two-dimensional polynomial functions. Both the film thickness and concentration were reconstructed using third-order polynomials in the variables $(S,R)$. For computational efficiency, the fitting procedure was performed in parallel over batches of pixels. Once calibrated, the fitted models were applied to the measured $(S,R)$ fields obtained during impact experiments, yielding spatially and temporally resolved maps of film thickness and concentration.

\begin{figure}
\centering
\includegraphics[scale=1]{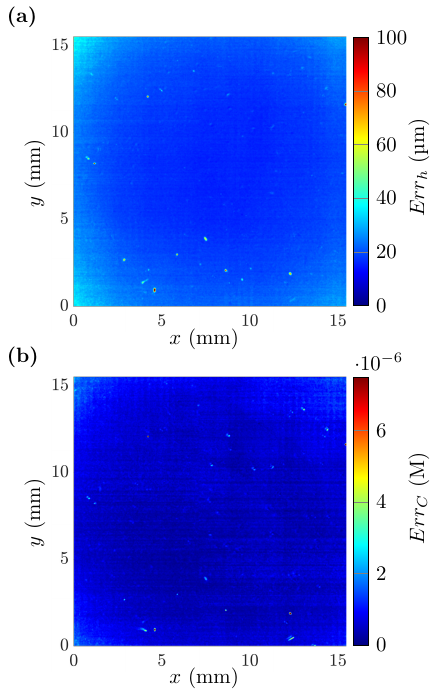}
\caption{Spatial distribution of reconstruction uncertainties obtained from the 2C-LIF calibration procedure. \textbf{(a)} film thickness reconstruction error $Err_h$ and \textbf{(b)} Rh6G concentration reconstruction error $Err_C$.}
    \label{fig:2}
\end{figure}

\begin{figure*}[b]
\begin{center}
\includegraphics[scale=0.78]{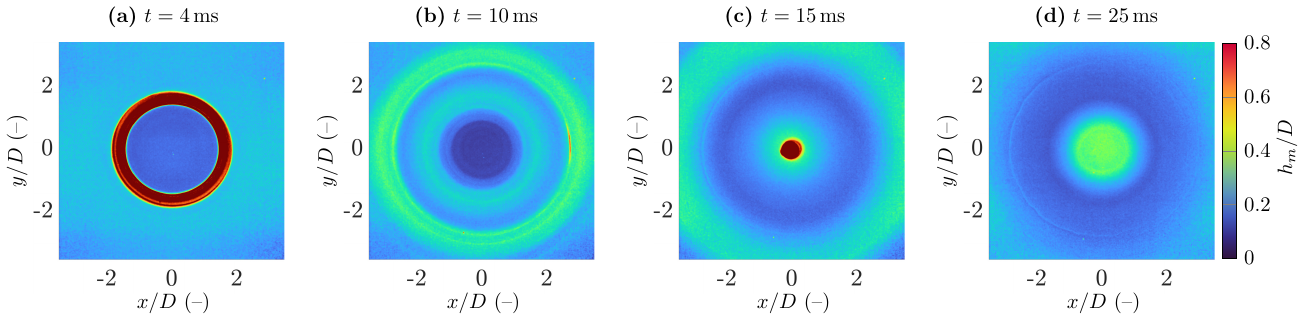}
\end{center}
\caption{Temporal evolution of the normalized film thickness field $h_m/D$ during droplet impact for $\delta=0.22$, $Re=4200$ and $We=110$. Snapshots at \textbf{(a)} $t=\SI{4}{\milli\second}$, \textbf{(b)} $t=\SI{10}{\milli\second}$, \textbf{(c)} $t=\SI{15}{\milli\second}$ and \textbf{(d)} $t=\SI{25}{\milli\second}$ show rim formation, cavity development, jet emergence and subsequent relaxation of the film. The color scale represents the reconstructed film thickness normalized by the initial droplet diameter, with warmer colors indicating thicker regions and cooler colors thinner regions.}
\label{fig:3} 
\end{figure*}

The calibration accuracy was validated with 68 independent measurements at known thicknesses and concentrations. A single validation measurement is performed by taking 100 images of thin liquid films with a know thickness and dye/ethanol concentration. At low reference concentrations, ratio-based error metrics are disproportionately influenced by small denominators which can exaggerate apparent deviations. To ensure an interpretable assessment of reconstruction performance across the entire concentration range, absolute error for each pixel was therefore evaluated. Similarly for thickness reconstruction, its error is calculated as
\begin{align}
	Err_{h}(x,y) &= |h_{m}(x,y)-h_{ref}(x,y)|, \label{eq:5}\\
	Err_{C}(x,y) &= |C_{m}(x,y)-C_{ref}(x,y)|, \label{eq:6}
\end{align}
where $h_m$ and $C_m$ denote the reconstructed local film thickness and concentration, respectively, and $h_{ref}$ and $C_{ref}$ the corresponding known reference thickness and concentration.

For water film, The spatial distribution of reconstruction accuracy was evaluated by calculating the mean absolute error for each pixel across all cases, separately for both film thickness and concentration fields (see Fig. \ref{fig:2}). In both reconstructions, the central region of the field of view exhibited the lowest errors, while slightly higher deviations occurred toward the image periphery. As mentioned previously, the increased error is attributed to vignetting effect, which reduces illumination intensity at the image edges, an effect particularly pronounced for low-magnification microscopy (\citealt{marty2007blank}). Overall, thickness errors remained predominantly below $\SI{30}{\micro\meter}$, whereas concentration errors were mostly below $2 \times 10^{-6}$ M. the lowest reconstruction errors were $\SI{15.88}{\micro\meter}$ for film thickness and $1.15 \times 10^{-6}$ M for dye concentration. 

Averaged over all pixels, the reconstructed error for water films was $\SI{24.18}{\micro\meter}$ in thickness and $1.67 \times 10^{-6}$ M in concentration. For ethanol–water films, slightly larger errors were observed, with mean errors of $\SI{57.36}{\micro\meter}$ in thickness and $2.24\%$ w/w in ethanol fraction. The decrease in accuracy for the ethanol case occurred predominantly for film thicknesses below $\SI{400}{\micro\meter}$, where partial evaporation induced Marangoni flows that caused film thickness to fluctuate strongly, as indicated by the confocal sensor during calibration. For water films, the errors were comparable to those reported in single-color LIF study (\citealt{ennayar2023lif}). The advantage of the present two-color LIF method is the ability to extract thickness and concentration simultaneously from a single measurement, which reduces experimental effort and avoids additional errors associated with sequential acquisition.

A key advantage of the 2C-LIF method was the ability to extract both thickness and concentration from a single measurement sequence, thereby reducing experimental effort and minimizing propagation of systematic error. In conventional single-color LIF, used in our previous study (\citealt{ennayar2023lif}), fluorescence signal provides only one observable, meaning that either local dye concentration or film thickness can be obtained, but not both simultaneously. As a result, two separate droplet impact experiments were required. One performed with fixed dye concentration to determine the film height evolution during the impact, and then another conducted to determine concentration using previously measured thickness. While in each step of this approach introduced additional uncertainty, the use of 2C-LIF eliminated this sequential dependency.

\section{Results and discussion}
\label{sec:4}
This section presents the application of the 2C-LIF technique to droplet impact on thin liquid films, with emphasis on species transport during impact. The reconstructed film thickness evolution is first shown to demonstrate that transient film deformation can be resolved concentration can be resolved under impact conditions. Subsequently, the capability of the method to resolve species transport is demonstrated through reconstruction of spatially and temporally resolved concentration fields. Mixing dynamics are quantified using the coefficient of variation of the concentration field, which provides a measure of homogenization during the impact. Finally, experiments with ethanol–water mixtures are presented to illustrate the applicability of the approach to multi-component liquids.

\subsection{Film thickness reconstruction during impact}
\label{sec:4.1}
The evolution of the liquid film thickness during droplet impact is examined in the following. Fig. \ref{fig:3} shows the temporal evolution of the reconstructed two-dimensional film thickness field for a representative case $\delta=0.22$, $Re = 4200$ and $We = 110$. The spatial coordinates $x$ and $y$, as well as the reconstructed film thickness $h_m$, are normalized by the initial droplet diameter $D$. The sequence illustrates the transient deformation of the liquid film following droplet impact.

\begin{figure}
\begin{center}
\includegraphics[scale=1.00]{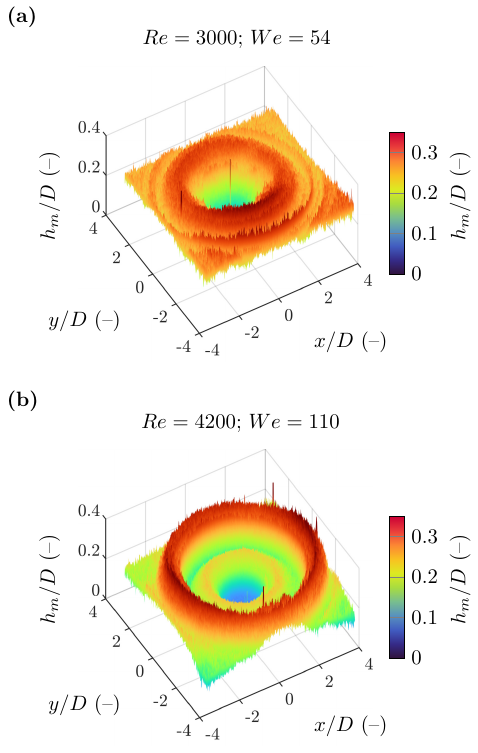}
\end{center}
\caption{3D reconstructions of the normalized film thickness $h_m/D$ at $t=\SI{10}{\milli\second}$ for $\delta=0.22$ under two impact conditions: \textbf{(a)} $Re=3000$, $We=54$ and \textbf{(b)} $Re=4200$, $We=110$. The color scale represents the film thickness normalized by the initial droplet diameter, highlighting differences in rim elevation and overall deformation with increasing impact inertia.}
\label{fig:4} 
\end{figure}

At early times ($t=\SI{4}{\milli\second}$) a pronounced annular region of elevated thickness form around the impact center. This structure corresponds to the elevated circular rim generated from the radially expanding lamella. The interior region exhibits a reduction in thickness due to the formation of a central cavity. As the impact progresses ($t=\SI{10}{\milli\second}$), the cavity deepens and the rim propagates further radially, while its amplitude decreases. At $t=\SI{15}{\milli\second}$, a localized increase in film thickness appears in the impact center, indicating the formation of an upward-directed jet. This central elevation subsequently diminishes ($t=\SI{25}{\milli\second}$), reflecting the relaxation of the interface.

While Fig. \ref{fig:3} provides a planar view of the thickness distribution, Fig. \ref{fig:4} presents three-dimensional reconstructions of the film height at $t=\SI{10}{\milli\second}$ for two impact conditions with identical initial film thickness $\delta=0.22$ but different impact velocities. The three-dimensional representation highlights fine features that are less apparent in 2D maps, such as small-amplitude surface undulations. For $Re=3000$ and $We=54$ (see Fig. \ref{fig:4}a), the deformation is characterized by a moderately elevated rim. Moreover, radially outward-traveling capillary waves are visible as small-amplitude ripples on the film surface. The rim remains relatively broad and smooth indicating limited vertical acceleration of the displaced liquid. In contrast, for the higher velocity case ($Re = 4200$ and $We = 110$ in Fig. \ref{fig:4}b) a more pronounced rim is observed. The steeper gradients at the rim indicate the onset of liquid sheet formation and the development of a crown structure. This transition is consistent with the higher Weber number, for which inertial forces dominate over surface tension, promoting crown formation.

\begin{figure*}[b]
\begin{center}
\includegraphics[scale=0.78]{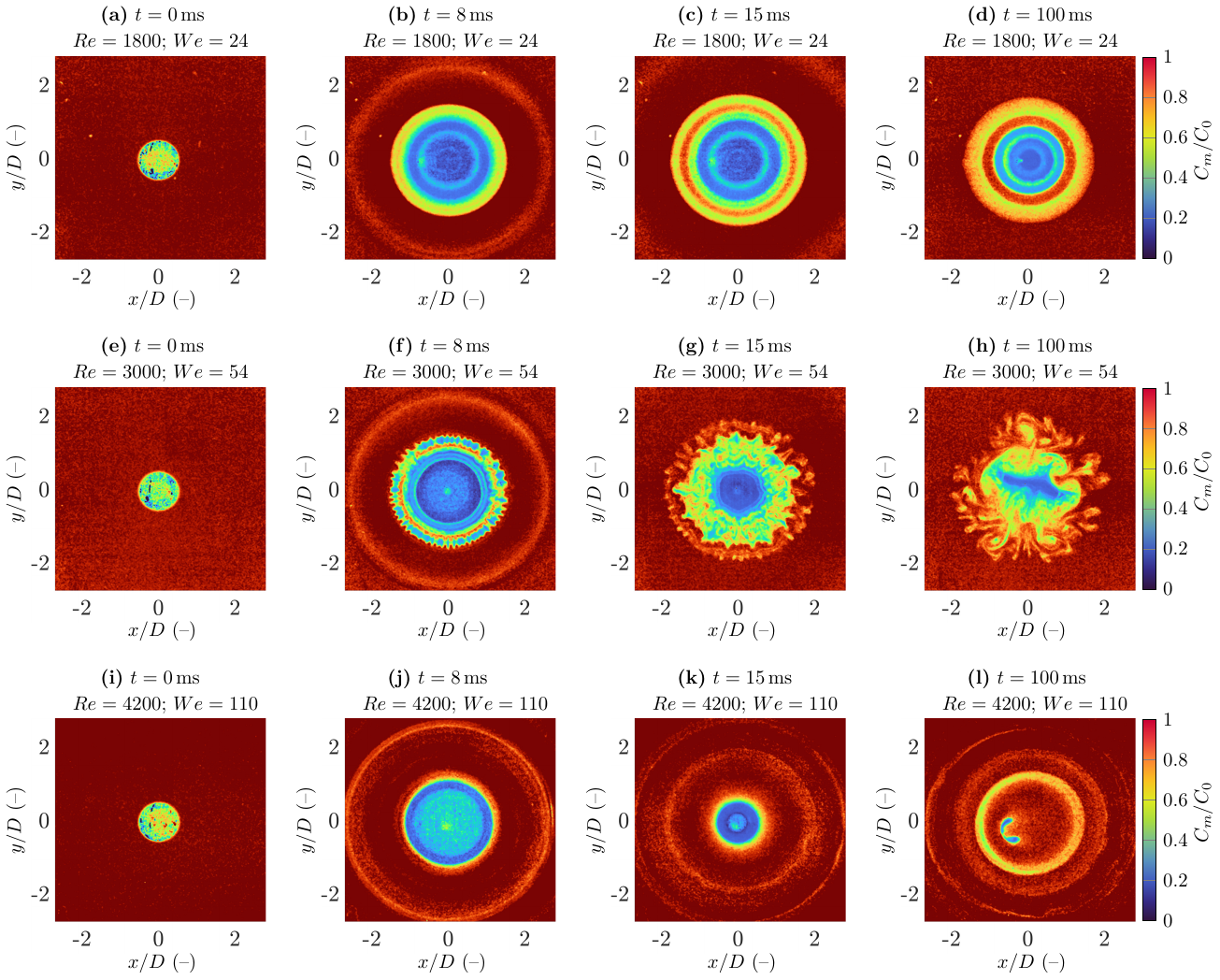}
\end{center}
\caption{Temporal evolution of the normalized concentration field $C_m/C_0$ during droplet impact on a liquid film with $\delta=0.22$. Rows corresponds to increasing impact velocity: \textbf{(a-d)} $Re=1800$, $We=24$; \textbf{(e-h)} $Re=3000$, $We=54$; \textbf{(i-l)} $Re=4200$, $We=110$. Columns show snapshots at $t=0$, $9$, $15$ and $=\SI{100}{\milli\second}$. The color bar indicates normalized concentration, where lower values correspond to droplet-rich regions.}
\label{fig:5} 
\end{figure*}

These results demonstrate that the 2C-LIF approach resolves film thickness variations with sufficient spatial and temporal resolution to capture key features of droplet impact, including cavity formation ripples propagation and crown evolution.

\subsection{Quantification of the species transport during impact}
\label{sec:4.2}
\begin{figure*}[b]
\begin{center}
\includegraphics[scale=0.78]{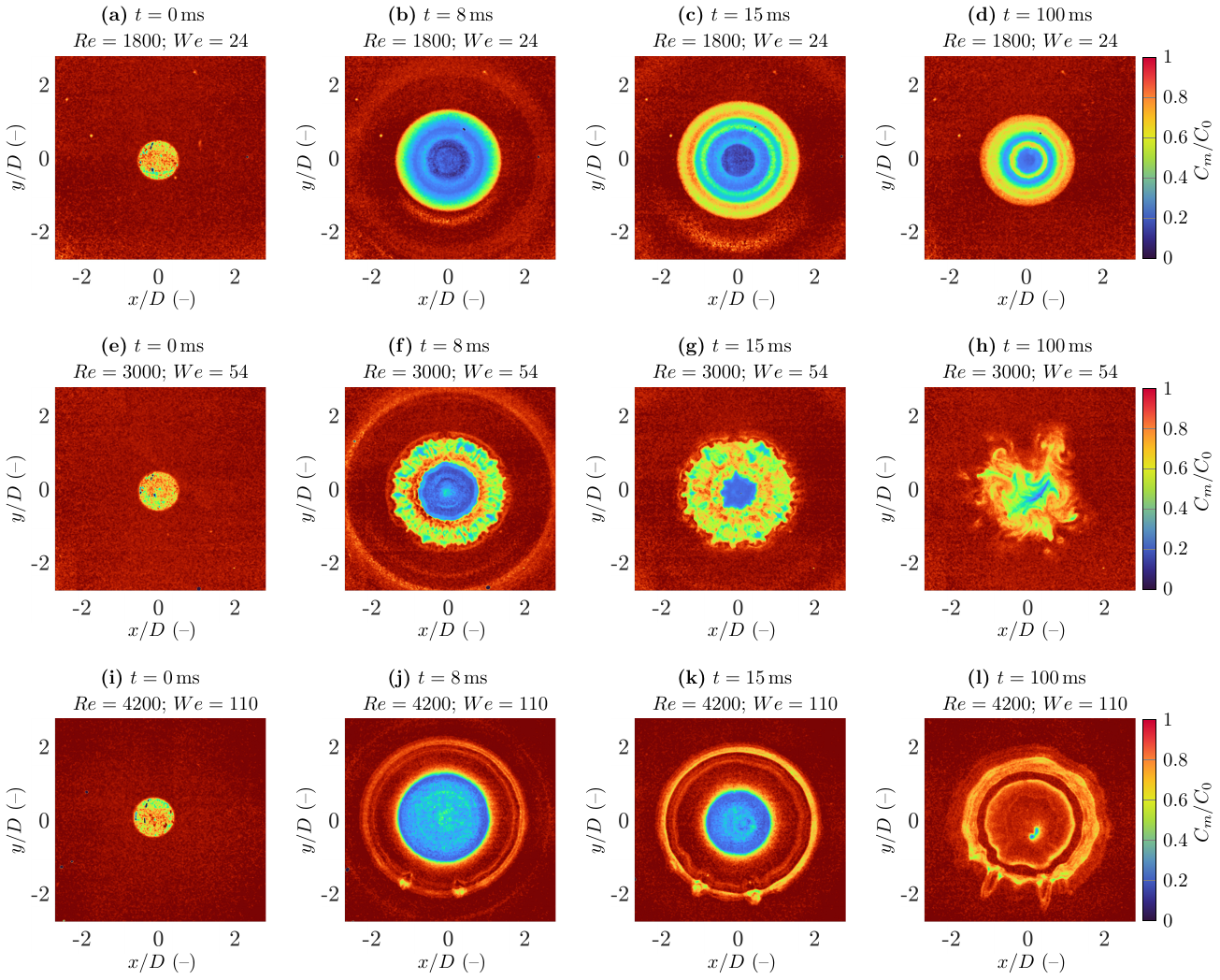}
\end{center}
\caption{Temporal evolution of the normalized concentration field $C_m/C_0$ during droplet impact on a liquid film with $\delta=0.36$. Rows corresponds to increasing impact velocity: \textbf{(a-d)} $Re=1800$, $We=24$; \textbf{(e-h)} $Re=3000$, $We=54$; \textbf{(i-l)} $Re=4200$, $We=110$. Columns show snapshots at $t=0$, $9$, $15$ and $=\SI{100}{\milli\second}$. The color bar indicates normalized concentration, where lower values correspond to droplet-rich regions.}
\label{fig:6} 
\end{figure*}

Figs. \ref{fig:5} and \ref{fig:6} show the spatiotemporal evolution of the reconstructed normalized concentration field $C_m/C_0$ during droplet impact for three velocities (corresponding to ``$Re=1800, We=24$"; ``$Re=3000, We=24$"; ``$Re=1800, We=24$") and two film thicknesses, $\delta=0.22$ and $\delta=0.36$, respectively. Here, $C_0$ denotes the initial Rh6G concentration in the undisturbed film, such that regions enriched with droplet liquid appear at lower values of $C_m/C_0$.

For the thinner film case $\delta=0.22$, the concentration evolution is first considered for the lowest impact velocity ($Re=1800$, $We=24$). The early-time concentration fields at $t=\SI{8}{\milli\second}$ (Fig. \ref{fig:5}b) and $t=\SI{15}{\milli\second}$ (Fig. \ref{fig:5}c) exhibit pronounced axisymmetric ring-shaped structures. This behavior is consistent vortex ring-driven transport mechanisms previously identified for impacts on finite-thickness films (\citealt{ennayar2025vortex}). In that work, the dominant mixing structures observed were shown to originate from formation and subsequent evolution of a droplet-induced vortex ring. The vortex ring forms due to azimuthal vorticity production at the liquid-liquid interface, driven by high velocity gradients that develop between the impacting droplet and the underlying liquid film (\citealt{cresswell1995drop}). The interaction of the vortex ring with the solid boundary strongly influences its later evolution and thereby the observed mixing structures. As shown in \citealt{ennayar2025vortex}, radial expansion of the vortex ring after approaching the wall produces initially axisymmetric ring-shaped mixing pattern. With decreasing film thickness or increasing vortex strength, associated with higher droplet impact velocity, boundary-layer separation during the radial expansion can generate a secondary vortex ring. This appears in bottom view as multiple concentric rings. The third observed regime is the occurrence of azimuthal perturbations that destabilize the coherent vortex rings structures and lead to their breakdown.

In agreement with this interpretation, the concentration fields obtained here for $\delta=0.22$, at $Re=1800$ and $We=24$ correspond to the multiple vortex rings regime identified in \citealt{ennayar2025vortex}. With increasing time, this structure evolves into multiple concentric rings with alternating concentration levels, which are clearly visible at $t=\SI{100}{\milli\second}$ (see Fig. \ref{fig:5}d). These rings reflect the the spiral motion associated with the primary vortex ring and the wall-induced secondary vortex ring, which entrain droplet and film liquid in different proportions. As a result, distinct radial zones form in which local mixture fraction varies, giving rise to the observed alternating concentration pattern. 

For the case $Re=3000$ and $We=54$ illustrated in Figs.~\ref{fig:5}e–h, azimuthal perturbations become visible at $t=\SI{8}{\milli\second}$ in the form of a circumferential fingering pattern. These perturbations indicate the onset of vortex-ring destabilization, which is likely associated with compression of the ring during the receding phase of the central cavity (\citealt{ennayar2025vortex}). As a consequence, the initially axisymmetric concentration field progressively loses symmetry and develops pronounced non-axisymmetric structures.

At the highest inertia case ($Re=4200$ and $We=110$), presented in Figs. \ref{fig:5}i-l, the evolution differs from the $We=54$ case. The concentration field initially forms a circular pattern, whereas azimuthal perturbations are not apparent at comparable times. This observation is noteworthy because the regime map in \citealt{ennayar2025vortex} was restricted to $We<64$. In the present case, the concentration fields indicate that droplet-rich liquid (low $C_m/C_0$) becomes strongly coupled to the central upward motion associated with jet formation. A related mechanism has been reported for droplet impact on deep liquid pools, where for $We>64$ the primary vortex ring is rapidly advected by the rising jet (\citealt{lee2015origin}). Although the present configuration involves a thin liquid film rather than a deep pool, the observed concentration redistribution suggests that a similar jet–vortex coupling may contribute to the mixing dynamics at these higher Weber numbers.

Increasing the initial film thickness to $\delta=0.36$, as shown in Fig.~\ref{fig:6}, leads to qualitatively similar mixing structures compared to the thinner film case, but with noticeable differences in the spatial extent of mixing. In particular, the radial region affected by concentration change is reduced relative to $\delta=0.22$. The thicker film delays the influence of the wall on the evolving vortex ring, thereby limiting its radial expansion, which in turn confines the spatial extent of mixing.

\begin{figure}
\begin{center}
\includegraphics[scale=0.91]{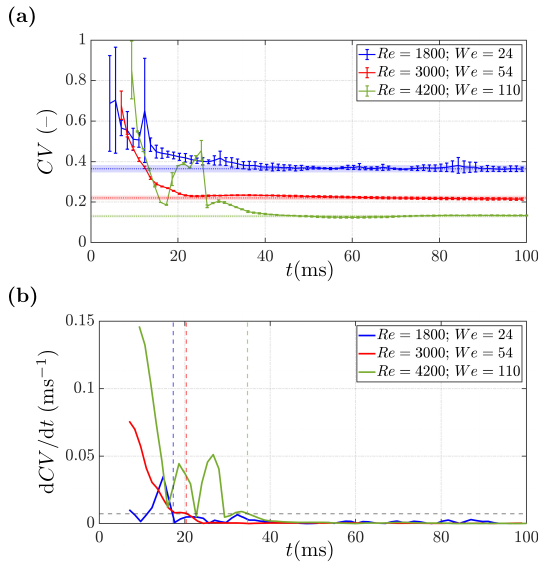}
\end{center}
\caption{Quantification of mixing dynamics for impacts on film of thickness $\delta=0.36$. \textbf{(a)} Temporal evolution of the coefficient of variation $CV$ for different impact conditions. The dashed horizontal lines denote the plateau values associated with diffusion-dominated mixing. \textbf{(b)} Absolute decay rate d$CV$/d$t$ as a function of time. The horizontal dashed line represents the threshold defined in Eq. (\ref{eq:8}), used to identify the onset of the diffusion-dominated regime. Vertical dashed lines mark the corresponding transition time $t_d$.}
\label{fig:7} 
\end{figure}

While the concentration fields discussed above provided insight into the mixing dynamics, a quantitative measure is required to assess the degree of homogenization during droplet impact. For this purpose, the coefficient of variation ($CV$) of the concentration field is used as a global indicator of mixing efficiency. In the context of scalar mixing, $CV$ quantifies spatial concentration inhomogeneity, with lower values corresponding to more homogeneous mixtures (\citealt{gandhi2011uv}). it is defined as
\begin{equation}
	CV(t) = \frac{\sqrt{\frac{1}{n-1}\displaystyle\sum_{x,y} \Bigr[C_{m}(x,y,t)-\frac{1}{n}\sum_{x,y} C_{m}(x,y,t)\Bigr]^2}}{\frac{1}{n}\displaystyle\sum_{x,y} C_{m}(x,y,t)},
	\label{eq:7}
\end{equation}
where $n$ denotes the number of pixels within the selected mixing region and $C_m(x,y,t)$ is the reconstructed local concentration at time $t$. The region of interest was defined based on the maximum lateral spreading of the impact and the detectable concentration gradient separating the actively mixed region from the surrounding film areas that retain their initial concentration.

Fig. \ref{fig:7}a shows the temporal evolution of $CV(t)$ for the investigated Reynolds–Weber number combinations at $\delta=0.36$, while Fig.~\ref{fig:7}b presents the corresponding decay rate d$CV$/d$t$, which provides a measure of the instantaneous homogenization rate. The $CV$ evaluation does not begin at $t=0~\mathrm{ms}$ because, during the very early stages of impact, strong optical reflections occur in regions with steep surface slopes, producing artificially elevated fluorescence intensities and unreliable concentration reconstruction. This limitation of LIF-based measurements has been documented previously (\citealt{hann2016study, ennayar2023lif}). The analysis therefore begins once surface waves amplitudes have sufficiently decreased relative to their wavelength.

For all investigated conditions, $CV(t)$ decreases gradually with time, reflecting progressive homogenization of droplet and film liquid. The curves eventually approach a plateau, indicating the transition from convection-dominated mixing to a diffusion-controlled regime. The horizontal dashed lines in Fig.~\ref{fig:7}a mark the plateau values associated with this diffusion-dominated state. Differences between impact conditions are evident in the rate at which $CV$ decreases. For the intermediate case ($Re=3000$, $We=54$), the decline is faster than for the lower-velocity case, consistent with enhanced vortex-driven convective mixing associated with vortex ring breakdown. For the highest velocity ($Re=4200$, $We=110$) an even steeper initial decrease is observed, reflecting stronger inertial transport. The temporary increase in $CV$ observed at intermediate times is attributed to jet formation, which redistributes droplet-rich liquid and transiently increases concentration heterogeneity before further mixing occurs. Furthermore, despite the inherently unsteady and partially chaotic flow conditions during vortex breakdown, repeat measurements (five repetitions per case) showed only small standard deviations in $CV(t)$, confirming the reproducibility of the observed mixing trends.

To quantify the onset of diffusion-dominated mixing, the diffusion-dominated time $t_d$ was defined as the earliest time for which 
\begin{equation}
	|dCV/dt|(t) \leq p\cdot|dCV/dt|_{\text{max}},
	\label{eq:8}
\end{equation}
where $\left| \frac{dCV}{dt} \right|_{\text{max}}$ denotes the maximum absolute decay rate of the coefficient of variation. This condition must persist for a minimum duration of $\Delta t = \SI{10}{\milli\second}$ in order to suppress noise-induced threshold crossings and avoid premature identification of the transition. Sensitivity tests performed with $p=0.02$, $p=0.05$, and $p=0.10$ yielded the same qualitative trends across all cases. Therefore, $p=0.05$ was adopted for the present analysis. The horizontal gray dashed line in Fig.~\ref{fig:7}b represents the threshold defined by Eq.~(\ref{eq:8}), while the vertical dashed lines indicate the corresponding diffusion-dominated times $t_d$ for each impact condition. A clear trend is observed, as increasing velocity prolongs the period of convective, vortex-driven mixing and delays the transition to the diffusion-dominated state.

\begin{figure}
\begin{center}
\includegraphics[scale=1]{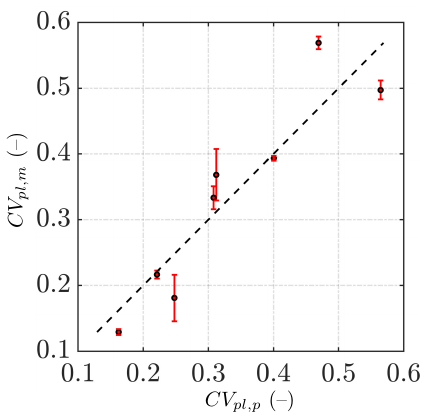}
\end{center}
\caption{Comparison between measured plateau coefficient of variation $CV_{pl,{m}}$ and values predicted from the empirical scaling relation (Eq.~\ref{eq:9}). The dashed line indicates ideal agreement.}
\label{fig:8} 
\end{figure}

The plateau value of the coefficient of variation $CV_{pl}$ provides a useful metric for comparing the extent of mixing achieved under different impact conditions and initial film thicknesses. Although complete homogenization is not reached within the experimentally accessible time window, the stabilized $CV$ level reflects the residual concentration variability after the dominant convective mixing processes have subsided. Examining how this plateau varies with Reynolds number and film thickness therefore allows a direct assessment of the influence of impact velocity and geometric confinement on the final degree of mixing. Such analysis facilitates comparison across impact regimes and provides a practical measure for characterizing mixing efficiency in droplet–film interaction studies.

The plateau coefficient of variation was analyzed using an empirical scaling relation. Regression of the measured data indicate that the apparent plateau value is described by
\begin{equation}
	CV_{pl}=A_0 Re^{-\alpha}\delta^{-\beta}+A_{\text{floor}},
	\label{eq:9}
\end{equation}
where $A_0=6.59$, $\alpha=0.0339$ and $\beta=0.0336$ are fitting parameters and $A_{\text{floor}}=-4.98$ represents a residual variability arising from measurement noise. The decrease of $CV_{pl}$ with increasing Reynolds number indicates enhanced advective homogenization at higher impact velocities. The inverse dependence on film thickness is attributed to the depth-averaged nature of the 2C-LIF measurement. Because the reconstructed concentration field represents an average over the local film thickness, identical droplet volumes produce larger normalized concentration variations in thinner films than in thicker ones.

Fig. \ref{fig:8} shows the agreement between the measured plateau values $CV_{pl,{m}}$ and their corresponding predictions from the empirical scaling $CV_{pl,{p}}$. The proposed relationship should be interpreted as a correlation valid only within the investigated parameter range rather than a universal scaling law, since additional parameters may further influence the behavior.

\subsection{Application to binary ethanol-water films}
\label{sec:4.3}
\begin{figure}[t]
\begin{center}
\includegraphics[scale=1]{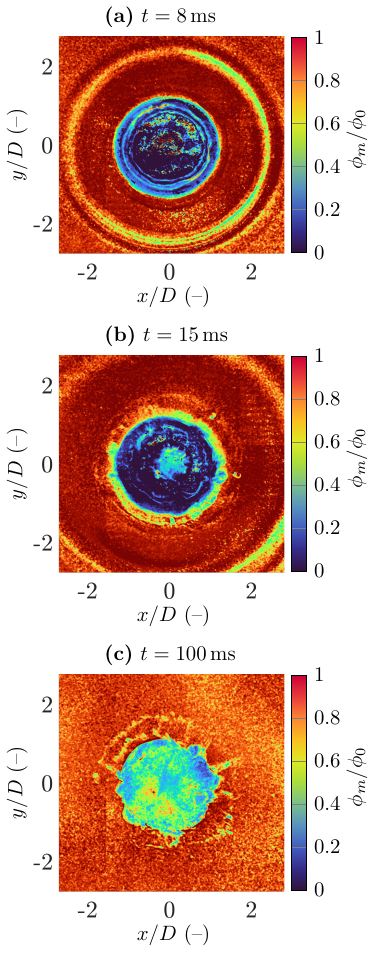}
\end{center}
\caption{Temporal evolution of the reconstructed normalized ethanol mass fraction field $\phi_m/\phi_0$ during the impact of a water droplet on a water-ethanol film ($30\%$ w/w ethanol) with $\delta=0.22$, $Re=1800$ and $We=24$. Snapshots at \textbf{(a)} $t=\SI{8}{\milli\second}$, \textbf{(b)} $t=\SI{15}{\milli\second}$ and \textbf{(c)} $t=\SI{100}{\milli\second}$ illustrate Marangoni-driven mixing. The color scale denotes the normalized ethanol mass fraction.}
\label{fig:9} 
\end{figure}

To demonstrate the broader applicability of the measurement approach, the analysis is extended to binary liquid films composed of water-ethanol mixtures. In contrast to single-component films, such systems introduce additional transport mechanisms arising from surface tension gradients and possible volatility effects, which can substantially influence the mixing dynamics during droplet impact. Examining these cases therefore provides a relevant test of the capability of the 2C-LIF technique for quantifying species transport in more complex liquid systems.

Fig. \ref{fig:9} presents reconstructed normalized ethanol mass fraction fields $\phi_m/\phi_0$ obtained from the 2C-LIF measurements for the impact of a water droplet onto a water-ethanol film containing 30\% ethanol at an initial thickness $\delta=0.22$ with $Re=1800$ and $We=24$. Here, the mixing dyamics differ markedly from the pure-water case observed in Sec. \ref{sec:4.2}. The presence of ethanol introduces surface tension gradient that generate Marangoni stresses along the interface. These stresses promotes additional small-scale convective motions, leading to mixing dynamics that differ markedly from those observed in single-component films. Under identical impact conditions, pure-water films exhibited coherent vortex ring structures governing the species transport, whereas in the ethanol-water film these later are strongly altered.

An additional observation is that the normalized ethanol mass fraction does not attain value of unity in regions outside the primary mixing zone. Two factors are considered responsible for this deviation. First, partial ethanol evaporation may occur during the time interval between film preparation and droplet impact, leading to a slight reduction of the local ethanol mass fraction and consequently affecting the reconstructed normalized values. Second, the 2C-LIF technique exhibits increased uncertainty for ethanol-water mixtures compared to water cases, as discussed in Sec. \ref{sec:3}.

\begin{figure}
\begin{center}
\includegraphics[scale=0.91]{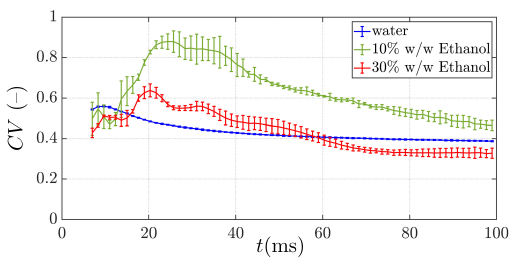}
\end{center}
\caption{Temporal evolution of the coefficient of variation $CV$ for droplet impact on pure-water and water-ethanol films with $10\%$ and $30\%$ w/w ethanol at $Re=1800$, $We=24$ and $\delta=0.22$.}
\label{fig:10} 
\end{figure}

Fig. \ref{fig:10} further illustrates the mixing behavior by presenting the temporal evolution of the coefficient of variation $CV$ for ethanol-water films with varying ethanol mass fractions at $Re=1800$ and $We=24$, for an initial film thickness of $\delta=0.22$. For the 10\% w/w ethanol film, strong surface tension gradients induce localized concentration variations that lead into an initial increase in $CV$. Subsequently, Marangoni-driven advection redistributes the liquid phases and progressively reduces concentration inhomogeneity. For the higher ethanol mass fraction (30\% w/w), a similar trend is observed but with overall lower $CV$ values. This reflects the larger initial ethanol content of the film, for which the same water-droplet volume produces comparatively smaller normalized compositional variations than the 10\% w/w case. Moreover, the presence of stronger solutal Marangoni stresses prolongs the convective mixing phase relative to the pure-water case. While water films transition earlier to diffusion-dominated regime, the ethanol-water-mixtures maintain surface tension-driven transport over longer times, resulting in sustained mixing and a continued larger decay rate of $CV$. 

Beyond these observations, the results demonstrate that the 2C-LIF approach developed in this work remains applicable to multi-component liquid configurations where additional effects, such as surface tension gradients, influence the mixing dynamics. 

\section{Conclusion}
\label{sec:5}
This study presented an experimental two-color laser induced fluorescence (2C-LIF) approach for the quantitative investigation of species transport during droplet impact on thin liquid films. While the method simultaneously reconstructs film thickness and scalar fields, the primary emphasis was placed on resolving and quantifying mixing dynamics with high spatial and temporal resolution.

For water films, the technique captured vortex-driven transport structures and their evolution across different Reynolds numbers, Weber numbers, and film thicknesses. Mixing was quantified using the coefficient of variation, enabling systematic comparison of homogenization between impact conditions and identification of the transition from convection-dominated to diffusion-dominated regimes. The plateau mixing level, determined from the asymptotic behavior of the coefficient of variation, was further correlated empirically with Reynolds number and film thickness within the investigated parameter range. Extension to water-ethanol films demonstrated that the approach remains applicable to multi-component systems, where solutal Marangoni stresses significantly alter and prolong the convective mixing process.

Overall, the main contribution of this work is the establishment of an experimental framework capable of directly quantifying species transport in thin liquid films upon droplet impact. This capability provides a foundation for future studies addressing complex systems and application-relevant impact scenarios.

\section*{Data availability statement}
Supplementary material can be found in. this contains all data plotted in figures, as well as the metadata of experiments. Recorded videos of droplet impacts, from which the images of figures are extracted, are also included

\section*{Acknowledgements}
%Projectnumber 237267381 –TRR 150
This project is funded by the Deutsche Forschungsgemeinschaft (DFG, German Research Foundation) – project number 237267381 – TRR 150, sub-project A07.


\bibliographystyle{apalike}
%\bibliography{LITERATURLISTE_exp}
%\bibliography{LITERATURLISTE_expRVW}
\bibliography{Bibliography_LIF_Drop}

\appendix       %Beginn des Anhangs

\end{document}